\documentclass[fleqn,12pt,twoside]{article}
\usepackage{espcrc1}

\usepackage{graphicx}
\usepackage[figuresright]{rotating}

\newcommand{\AmS}{{\protect\the\textfont2
  A\kern-.1667em\lower.5ex\hbox{M}\kern-.125emS}}

\title{Tests of the Polyakov Loops Model}

\author{Robert D. Pisarski
	\address{Department of Physics,\\Brookhaven National Laboratory,\\
	Upton, NY  11973 USA}
        \thanks{I thank the U.S. Dept. of Energy Contract 
DE-AC-02-98CH-10886 for their support}}
       
\begin{document}

\maketitle

\begin{abstract}
I give a brief review of the Polyakov Loops Model and tests thereof.
I concentrate especially on how in a pure
$SU(N)$ gauge theory, Polyakov loops with $Z(N)$ charges
two and three affect the effective potential for charge-one loops.  
\end{abstract}

\section{Introduction}

Recently I suggested that for the deconfining phase transition
in a pure gauge theory, the behavior of the usual order parameter ---
the Polyakov loop --- may be correlated with the pressure.  

In fact this correspondence is implicit in studies of two
colors done by Engels, Fingberg, Redlich, Satz, and Weber 
in 1989 \cite{su2a}.
For two colors, the deconfining transition appears to be of second
order \cite{su2a,su2b,su2c}.  
This means that the behavior of the free energy should be
singular as $T \rightarrow T_c$, where $T_c$ is the transition
temperature.  For ordinary spin systems, if $p$ is the pressure,
\begin{equation}
p(T) \sim (T_c - T)^{2 - \alpha}
\label{e1}
\end{equation}
with $\alpha$ the critical exponent for the Ising model in
three dimensions \cite{svetitsky}.

Now consider how this might happen for the deconfining transition
of two colors.  One must have singular behavior near $T_c$, (\ref{e1}).
Moreover, however, by asymptotic freedom the pressure must approach
the ideal gas law at infinite temperature, 
$p \sim T^4$ as $T \rightarrow \infty$.  The simplest guess which
satisfies both limits, $T \rightarrow T_c^+$ and $T\rightarrow \infty$, is
\begin{equation}
p(T) \sim (\Delta t)^{2 - \alpha} \; T^4 \;\;\; , \;\;\;
\Delta t = \left( 1 - \frac{T_c}{T} \right) \; .
\label{e2}
\end{equation}
Using this form, Engels {\it et al.} \cite{su2a} fit to the
pressure above $T_c$.  The fit is not perfect, and would hopefully
improve with improved Lattice actions.

While (\ref{e2}) appears inconsequential, it is not.
In mean field theory, (\ref{e1}) arises from a
potential, such as
\begin{equation}
{\cal V} \; = \; - \;
\frac{m_1^2}{2} \; \ell_1^2 \; + \; \frac{\lambda_1}{4} \; \ell_1^4
\label{e3}
\end{equation}
I haven't defined the field $\ell_1$ yet, but the potential 
is appropriate to 
the universality class of the Ising model,
given a real field invariant under the symmetry of
$\ell_1 \rightarrow - \ell_1$.
At the minimum of the potential, 
the vacuum expectation value is 
$\ell_0 = \langle \ell_1 \rangle = \sqrt{m_1^2/\lambda_1}$, and 
the potential is ${\cal V}(\ell_0) \sim \ell_0^4 \sim m_1^4$.
Taking $m_1^2 \sim T_c - T$, one obtains (\ref{e1}), at least in the mean
field limit, where $\alpha = 0$. 
To obtain (\ref{e2}), however, it is necessary
to multiply this potential times an {\it overall} factor of $T^4$.
The pressure is then given by $T^4$ times $- {\cal V}(\ell_0)$:
\begin{equation}
p \; = \; \frac{\ell_0^4}{4} \; T^4 \; .
\label{e4}
\end{equation}
Choosing $m_1^2 \sim T - T_c \sim 
\Delta t$ then agrees with (\ref{e2}), taking 
$\alpha = 0$.  Including critical effects, $\alpha \neq 0$,
should be doable, and doesn't
affect the novel, albeit non-universal, feature of ${\cal V}$, which is
having a potential times $T^4$.  Notice that to
obtain the correct limit at high temperature, one has to
define a reduced temperature as a function of $T_c/T$, and not as
$T/T_c$, as is customary for a spin model.  This is because a
spin model condenses at low temperatures, and so is smooth
as $T \rightarrow 0$, while gauge theories condense
only at high temperatures, and so are smooth
as $T \rightarrow \infty$.

The Polyakov Loops Model (PLM) \cite{plm1,plm2,plm3} attempts
to make this correspondence precise.  At a nonzero temperature
$T$, in Euclidean spacetime the thermal Wilson line is
\begin{equation}
{\bf L}(\vec{x}) \; = \; 
{\cal P} \exp \left( i g \int^{1/T}_0 A_0(\vec{x},\tau) \, 
d\tau \right) \; ,
\label{e6}
\end{equation}
$A_0$ is in the fundamental representation.
This transforms as an adjoint field under the local $SU(N)/Z(N)$ 
gauge symmetry, and as a field with charge one under the global
$Z(N)$ symmetry.  To obtain a gauge invariant operator, the
simplest thing to do is to take the trace, forming the usual Polyakov loop,
\begin{equation}
\ell_1 = \frac{1}{N} \; {\rm tr}\left( {\bf L} \right) \; .
\label{e7}
\end{equation}
This transforms under the global $Z(N)$ as a field with charge one.
The expectation value of $\ell_1$ is only nonzero above $T_c$,
which is the temperature for the deconfining phase transition.
It is the trace of the propagator for an infinitely massive (test) quark,
as the propagator is just the color Aharanov-Bohm phase factor.

In fact $\ell_1$ is only the first in an infinite series of 
gauge-invariant operators.  For example, consider 
\begin{equation}
\ell_2 = \frac{1}{N} {\rm tr}\left( {\bf L}^2 \right) - \ell_1^2 \; .
\label{e8}
\end{equation}
Under the global $Z(N)$ symmetry,
this Polyakov loop has charge two.
I subtract off $\ell_1^2$ 
to obtain an independent field.  The field $\ell_2$
is the trace of the wave function for two quarks, sitting on top
of each other, minus the square of the trace for each propagator.

The series continues indefinitely.  One term of especial interest
is that for the $Z(N)$ singlet $\ell_N$,
\begin{equation}
\ell_N = \frac{1}{N} \; {\rm tr}( {\bf L}^N ) - \ldots
\end{equation}
This is a ``quarkless baryon'', as the trace of $N$ propagators,
with the pieces from other traces subtracted off.
Unlike any Polyakov loop with lower $Z(N)$ charge, the baryon loop
has a nonzero vacuum expectation value at any temperature, including zero.

As matrix traces of phase factors, 
Polyakov loops are dimensionless numbers.  
The basic Polyakov loop, $\ell_1$, is naively bounded by one.
In an aymptotically free gauge theory without dynamical quarks,
by choice of renormalization condition one should be able to fix that
$\ell_0\rightarrow 1$ at $T=\infty$.
The expectation values of
Polyakov loops with higher charge $<N$ vanish below $T_c$, grow, and then
vanish at infinite temperature, as the 
expectation value of powers of $g A_0$.
As a singlet, expectation value of the baryon loop
$\ell_N$ is always nonzero, even at $T=0$.

In original work \cite{plm1,plm2}, only the charge-one
Polyakov loop was considered.  For four or more colors,
the importance of Polyakov loops with higher charges 
was considered in \cite{plm3}.
In this note I discuss what role they play for two, three, and four colors,
without dynamical quarks.

All that I will discuss are elementary exercises
in mean field theory, albeit with a slightly peculiar potential,
and a number of effective fields which grows with the number of colors.
In all cases I assume that the only critical field is 
the charge-one loop, $\ell_1$, so the classification of universality
classes is standard \cite{svetitsky}.  I consider how non-critical
fields may alter non-universal properties of the deconfining phase
transition.  I use the renormalization group,
concentrating on renormalizable couplings, starting with those
with the greatest mass dimension.  
Numerous predictions are directly testable by Lattice simulations.

\section{Two Colors}

For two colors,
consider the charge-one and two Polyakov
loops.  In $SU(2)$, the trace of any element of the Lie group is
real, so both are real fields.  
The charge-one loop $\ell_1$
is the trace of the Wilson line in the fundamental
representation; the charge-two, or baryon, loop $\ell_2$ 
is the trace in the adjoint representation.

The charge-one loop changes sign under a global $Z(2)$ transformation, 
$\ell_1 \rightarrow - \ell_1$, but as a singlet, 
the baryon loop does not, $\ell_2 \rightarrow + \ell_2$.
Besides the potential for charge-one loops in (\ref{e3}),
the potential for the baryon loop starts as:
\begin{equation}
{\cal V}_2 \; = \; h \; \ell_2 \; + \; \frac{1}{2} \; m_2^2 \; \ell_2^2
\; + \; \lambda_2 \; \ell_1^2 \; \ell_2  + \ldots \; .
\label{e9}
\end{equation}
To lowest order, there is a term linear in $\ell_2$, a mass
term, and a cubic coupling to the charge-one loop.
General terms are powers of $\ell_2$ times powers of $\sim \ell_1^2$.
To lowest order, the expectation value of the baryon loop is
\begin{equation}
\langle \ell_2 \rangle \; = \; - \; \frac{h + \lambda_2 \, 
\ell_0^2}{m_2^2} \; .
\label{e10}
\end{equation}
As expected, the linear term implies $\langle \ell_2 \rangle \neq 0$
at all temperatures.

The Lattice gives us useful information about these coupling 
constants \cite{su2a,su2b,su2c}.
The charge-one field behaves as expected, zero below $T_c$,
nonzero above, in the universality class of the Ising model.
What is unexpected, and first stressed by Damgaard \cite{su2b},
is that $\ell_2$ behaves {\it almost} like $\ell_1$: 
its expectation value is very small --- but nonzero --- below $T_c$, and
then grows sharply above $T_c$.
That it is small below $T_c$ implies that the coefficient
of the linear term is small, $h\approx 0$.  Because its
expectation value grows with $\ell_1$, the
baryon loop couples strongly to $\ell_1$, with $\lambda_2 \gg 1$.
This assumes that the mass for the baryon loop, $m_2^2$, 
stays large and positive as the mass of the charge-one loop, $m_1^2$,
changes sign at $T_c$.  This is the simplest possibility.

The pressure is also small below $T_c$.  This, and the small
expectation value of $\ell_2$ at $T < T_c$, 
is consistent with the assumption of the PLM,
that condensates for Polyakov loops generate pressure.
After all, if a potential for $\ell_1$ is multiplied by $T^4$,
then consistency demands the same for every potential involving
any Polyakov loop, whatever the charge.  
Thus a significant expectation value for
$\ell_2$ below $T_c$, 
with no corresponding pressure, would contradict the PLM.

Conversely, the condensate for the baryon loop contributes,
through its potential, to the pressure above $T_c$.  
As stated, I assume that the mass
of $\ell_2$ is large and positive.  Then $\ell_2$ can be integrated out,
giving:
\begin{equation}
{\cal V}_{\rm eff} \; \sim \; 
- \frac{\lambda_2^2}{m_2^2} \; \ell_1^4 \; .
\label{e10a}
\end{equation}
The interesting feature of (\ref{e10a}) is the sign; with a real
coupling of $\ell_2$ to $\ell_1^2$, integrating $\ell_2$ out necessarily 
generates a quartic coupling of $\ell_1$ which is {\it negative}.  
For two colors, this doesn't seem to be of consequence, as the
transition is of second order.  This requires that
the the total quartic coupling for $\ell_1$ is 
positive, 
\begin{equation}
\widetilde{\lambda}_1
\; = \; 
\lambda_1 - \frac{\lambda_2^2}{m_2^2} \;  > \; 0
\label{e10b}
\end{equation}
For four or more colors,
however, the total quartic coupling is negative,
$\widetilde{\lambda}_1 < 0$.

The suceptibility for each loop is also interesting.
As expected at a critical point,
the susceptibility for $\ell_1$ diverges.
If $\ell_2$ is not a critical
field, and does not couple to $\ell_1$,
then its susceptibility is well behaved about $T_c$.  Hwever, in
the PLM, $\ell_2$ couples to $\ell_1$, and thus at the critical point
its susceptibility should approach $\lambda_2^2$ times that
for $\ell_1^2$.  
This appears to disagree with present Lattice data \cite{su2b},
which finds a finite susceptibility for $\ell_2$.

\section{Three Colors}

\subsection{Baryon Loop}

The morale of the deconfining transition for two colors is
promising.  While in principle the baryon loop could have
had a large expectation value below $T_c$, it does not.  Like
the pressure below $T_c$, it is there, but is so small that it
almost doesn't matter.  This means that the potential for
the baryon loop is dominated by the coupling to charge-one loops.

Let us assume that this behavior holds for any number of colors, $N \geq 3$.
As for two colors, we assume that any term linear in $\ell_N$ is small.
Further, that the potential for the baryon loop is dominated by
the coupling to the charge-one loop.  The simplest possible coupling is
\begin{equation}
\lambda_N \; \ell_N^* \; (\ell_1)^N  + {\rm c.c.}\; .
\label{e12}
\end{equation}
Perhaps $\ell_1$ dominates because it is like a dilaton,
and couples in a universal, gravitational manner to all fields.
Given the experience with two colors,
however, this is the natural coupling to assume.  Then the
expectation value of the baryon loop grows like the
$N^{\rm th}$ power of $\langle \ell_1\rangle$;
how the baryon loop condenses is given 
in terms of a single coupling constant.

Thus it is
{\it much} easier making baryon anti-baryon pairs
above the deconfining transition.  This effect turns on more
strongly as the number of colors increases.  
For three colors, each baryon costs $\ell_0^3$, or $\ell_0^6$
for a baryon anti-baryon pair.  

Notice, however, that the baryon loop doesn't significantly affect the
pressure above $T_c$ for three or more colors.  Integrating $\ell_N$
out, if the coupling in (\ref{e12}) is dominant, this gives a term
as in (\ref{e10a}):
\begin{equation}
{\cal V}_{\rm eff} \; \sim \; 
- \frac{\lambda_N^2}{m_2^2} \; (|\ell_1|^2)^N \; .
\label{e12a}
\end{equation}
For three colors this is a six-point interaction.  The total coupling
$(|\ell_1|^2)^3$ must be positive, which requires 
$\widetilde{\lambda}_6 = \lambda_6- \lambda_N^2/m_2^2>0$.  
As the total six-point interaction must be positive for stability,
$\widetilde{\lambda}_6>0$, this is not a very useful constraint.

\subsection{Charge-two Loop}

Now I consider the charge-two loop, $\ell_2$, and its coupling to the
charge-one loop, $\ell_1$, for the specific case of three colors.
As the $Z(3)$ group is cyclic, the charge-two loop is equivalent
to a loop with charge minus one.  This is only in terms of the
$Z(3)$ charge; it is an independent field, and both must
be included in an effective lagrangian.

For three (or more) colors, all $\ell_j$ are complex numbers.  
The couplings include those only involving $\ell_1$,
\begin{equation}
{\cal V}_1 \; = \; - \;
\frac{m_1^2}{2} \; |\ell_1|^2 \; + 
\; \frac{\kappa_1}{3} \; (\ell_1^3 + {\rm c.c.}) \; + 
\; \frac{\lambda_1}{4} \; (|\ell_1|^2)^2
\label{e13}
\end{equation}
those only involving $\ell_2$,
\begin{equation}
{\cal V}_2 \; = \; + \;
\frac{m_2^2}{2} \; |\ell_2|^2 \; + 
\; \frac{\kappa_2}{3} \; (\ell_2^3 + {\rm c.c.}) \; + 
\; \frac{\lambda_2}{4} \; (|\ell_2|^2)^2
\label{e14}
\end{equation}
and cross-terms which mix the two:
\begin{equation}
{\cal V}_3 \; = \; 
\zeta_1 \; (\ell_1 \, \ell_2 + {\rm c.c.}) \;
+ \; \zeta_2 \; |\ell_1 \, \ell_2|^2
+ \; \zeta_3 \; ( (\ell_1 \, \ell_2)^2 + {\rm c.c.}) \; .
\label{e15}
\end{equation}

The term involving only the charge-one loop $\ell_1$, (\ref{e13}), 
is customary \cite{svetitsky}.
Because of the cubic $Z(3)$ invariant coupling $\kappa_1$,
the transition is of first order.
I assume that the mass for the
charge-one loop changes sign at $T_c$, but that the mass for
the charge-two loop is large, positive, and approximately constant.  
(There are other reasons: if $m_2^2$ changed
sign, the local $SU(N)$ symmetry could break, which is
not expected.)

Nevertheless, it is not true that the charge-two loop can be neglected.
It {\it directly} mixes with the charge-one loop through a coupling
$\zeta_1 \neq 0$, (\ref{e15}).  If this coupling is nonzero, and
the charge-two field is light, then it could affect the transition
signficantly.  In particular, condensates for the charge-two loop do change the
pressure above $T_c$.  

However, by numerical simulations it would be easy to directly
measure both the mass of the charge-two loop, and its coupling to
the charge-one loop.  One thing which is know from the Lattice is that the
deconfining transition is relatively weakly first order: the ratio
of the latent heat, to the ideal gas energy, is approximately 
$\sim 1/3$ \cite{su3}.
This is in contrast to a bag model, where this ratio is $=4/3$.
A weakly first order transition
requires that the cubic coupling for $\ell_1$, $\kappa_1$
in (\ref{e13}), is small.  Both this cubic coupling, and
the mixing term between $\ell_1$ and $\ell_2$, are terms which
are invariant under the $Z(3)$ symmetry, but not under the
$U(1)$ symmetry.  
In terms of the Wilson line, $\ell_1^3 \sim ({\rm tr}{\bf L})^3$,
and $\ell_1 \ell_2 \sim ({\rm tr}{\bf L})({\rm tr}{\bf L}^2)$.
All other terms in the potential are $U(1)$ invariant.
So maybe {\it all} of the couplings invariant under $Z(3)$, but
not $U(1)$, are small.
If so, the charge-one loop couples weakly to the charge-two loop,
and the charge-two loop can be neglected.  

Naive analysis indicates that after
integrating $\ell_2$ out, one obtains a new effective potential
involving only $\ell_1$, similar in form to 
(\ref{e13}).  This suggests that even if $\ell_2$ mixes strongly
with $\ell_1$, its effects are relatively innocuous.

\section{Four Colors}

For four colors, I consider the Polyakov loops with charge one, $\ell_1$,
and two, $\ell_2$.  
Polyakov loops with charge three are equivalent to charge minus one;
this presumably is like charge minus one loops for three colors.
By previous argument, after integrating out the charge-four, or baryon, loop,
one obtains a contribution to the pressure of eighth order, which I
neglect as non-renormalizable.

Under a global $Z(4)$ transformation,
\begin{equation}
\ell_1 \; \rightarrow \; i \, \ell_1 \;\;\; , \;\;\;
\ell_2 \; \rightarrow \; - \, \ell_2 \; .
\label{e16}
\end{equation}
The potential for $\ell_1$ starts as
\begin{equation}
{\cal V}_1 \; = \; - \;
\frac{m_1^2}{2} \; |\ell_1|^2 \; + 
\; \frac{\lambda_1}{4} \; (|\ell_1|^2)^2 \;
+ \; \frac{\kappa_1}{4} \; (\ell_1^4 + {\rm c.c.}) \; .
\label{e17}
\end{equation}
The potential for $\ell_2$ is like that for the charge-one field of
two colors, (\ref{e9}):
\begin{equation}
{\cal V}_2 \; = \; + \;
\frac{m_2^2}{2} \; |\ell_2|^2 \; + 
\; \frac{\lambda_2}{4} \; (|\ell_2|^2)^2 \;
+ \; \frac{\kappa_2}{4} \; (\ell_2^4 + {\rm c.c.}) \; .
\label{e18}
\end{equation}
Lastly, there is a mixing term,
\begin{equation}
{\cal V}_3 \; = \; 
\zeta_1 \; ( \ell_2^* \, \ell_1^2 + {\rm c.c.}) \;
+ \; \zeta_2 \; ( \ell_2 \, \ell_1^2 + {\rm c.c.}) \; .
\label{e19}
\end{equation}
As for two colors, (\ref{e10a}), after integrating $\ell_2$ out, 
and taking $\zeta = \zeta_1 = \zeta_2$ for simplicity, there is a negative
quartic coupling for $\ell_1$, $\sim \zeta^2/m_2^2$.

In general, a negative quartic coupling constant indicates a first
order transition; to ensure stability for large fields,
the six-point coupling constant must be positive.  This 
assumes renormalizable couplings in three dimensions, as appropriate
to the finite temperature transition in four space-time dimensions.
Thus if the coupling constant(s) $\zeta$ are large enough, the
charge-two loop can drive the transition --- which otherwise would
be second order, for $\lambda_1 > 0$ --- first order.

This is what present Lattice results find \cite{su4}.  The ratio
of the latent heat, to the ideal energy density at $T_c$, is approximately
$2/3$, versus $1/3$ for three colors.  This can be tested on the
Lattice by measuring the couplings between the charge-one and two loops,
$\sim \zeta$, which should be large.  

\section{Five or More Colors}

Five colors should be like four.  The charge one and minus one
loops probably behave like three colors, mixing linearly,
perhaps with small coefficient.  The charge two (equals charge minus three)
field can couple with $\zeta_1 \neq 0$, as in (\ref{e19}), while
$\zeta_2 = 0$ because of the $Z(5)$ invariance.  Even so,
$\zeta_1 \neq 0$ can drive the transition first order, as for four colors.

However, the tendency does not continue.  For six colors, consider
the field with charge three, $\ell_3$; it couples to that with
charge one through two quartic couplings,
\begin{equation}
{\cal V}_3 \; = \; 
\zeta_1 (\ell_3^* \ell_1^3 + {\rm c.c.}) \; .
+ \; \zeta_2 ( \ell_3 \ell_1^3 + {\rm c.c.}) \; .
\label{e21}
\end{equation}
Integrating out $\ell_3$, this generates a six-point interaction
between the $\ell_1$'s,
\begin{equation}
{\cal V}_{\rm eff} \; \sim \; 
- \frac{\zeta^2}{m_3^2} \; (|\ell_1|^2)^3 \; ;
\label{e22}
\end{equation}
again, this constraint is not useful, as the total six-point interaction
must be positive.

Similarly, for eight colors, the field with charge four generates
an eight-point interaction between $\ell_1$'s, which is an
irrelevant operator.  Thus for six or more colors, the charge-two
loop uniquely drives the transition first order.  

How strongly the transition is of first order is not known from
the mean field analysis of the PLM.  Normalizing the color traces
so that Polyakov loops are naturally of order one, presumably the
overall potential for Polyakov loops, whatever their charge,
is multiplied by an overall factor of $\sim N^2$.  Thus the
transition can be strongly first order in the large $N$ limit.

\section{Renormalization}

Classically, the charge-one Polyakov loop is a complex number
bounded by one.  Quantum mechanically, it is a renormalizable
(composite, nonlocal) operator.
Different traces of powers of the Wilson line have
different renormalization constants.  The charge-one loop $\ell_1$
can be fixed to have unit magnitude at infinite temperature.  
For the charge-two loop, one needs to introduce
seperate renormalization constants for both ${\rm tr}{\bf L}^2/N$
and $\ell_1^2$.  Again, each is required to approach one at
$T = \infty$.  The renormalized $\ell_2$ is the difference
of these two renormalized operators.  

How to do this in a practical sense with a lattice cutoff is
obscure at present.  A different approach has been investigated
in \cite{renorm}, by matching to the potential at short distances.

\section{Dynamical Quarks}

I conclude with an interesting example how the PLM might apply
with dynamical quarks.  Digal, Laermann, and Satz \cite{ds} have studied
the three color theory with two light flavors of dynamical quarks.
The pressure below $T_c$ from three pions is small compared to
the pressure of quarks and gluons above $T_c$, so the PLM appears
to work according to ``flavor independence'' \cite{flind}.

The susceptibilities for the chiral order parameter, and for the 
(charge-one) Polyakov loop, were measured about a transition
which is apparently a second order, chiral transition.  
The susceptibility for the
chiral order parameter grows in a manner expected for a critical
field (presumably in the $O(4)$ universality class).  
Further, on present lattices,
the susceptibilities for the Polyakov loop (with
respect to temperature or the chiral order parameter), also appear
to grow, in fact even stronger than that for the chiral order
parameter.  While at first
this appears surprising, if there is a cubic coupling between
the charge-one loop and two chiral fields, $\Phi$,
$\sim \kappa \ell_1 {\rm tr}\Phi^\dagger \Phi$, then 
this might explain the divergences for the 
susceptibilities of the Polaykov loop \cite{rajagopal}.
However, such a coupling generates an expectation value for
$\ell_1$ below $T_c$, where it does not appear to be large.

\end{document}